\documentclass[aip,jap,reprint,showpacs,floatfix]{revtex4-1}

\usepackage{graphicx,color}
\usepackage{amsmath,amssymb}
\pdfoutput=1

\newcommand{\Nmax}{%
  \ensuremath{N_{\mathrm{max}}}}
\newcommand{\Ibias}{%
  \ensuremath{I_{\mathrm{bias}}}}
\newcommand{\jbias}{%
  \ensuremath{j_{\mathrm{bias}}}}
\newcommand{\Rhc}{%
  \ensuremath{R_{hc}}}
\newcommand{\lGL}{%
  \ensuremath{\lambda_{\mathrm{GL}}}}
\newcommand{\Idep}{%
  \ensuremath{I_{c,\mathrm{dep}}}}

\begin{document}

\title{Numerical analysis of detection-mechanism models of SNSPD}

\author{Andreas Engel} \email[]{andreas.engel@physik.uzh.ch}
\author{Andreas Schilling}
\affiliation{Physics Institute of the University of Zurich, Winterthurerstr.\ 190, 8057 Zurich, Switzerland}

\date{\today}

\begin{abstract}
The microscopic mechanism of photon detection in superconducting nanowire single-photon detectors is still under debate. We present a simple, but powerful theoretical model that allows us to identify essential differences between competing detection mechanisms. The model is based on quasi-particle multiplication and diffusion after the absorption of a photon. We then use the calculated spatial and temporal evolution of this quasi-particle cloud to determine detection criteria of three distinct detection mechanisms, based on the formation of a normal conducting spot, the reduction of the effective depairing critical current below the bias current and a vortex-crossing scenario, respectively. All our calculations as well as a comparison to experimental data strongly support the vortex-crossing detection mechanism by which vortices and antivortices enter the superconducting strip from the edges and subsequently traverse it thereby triggering the detectable normal conducting domain. These results may therefore help to reveal the microscopic mechanism responsible for the detection of photons in superconducting nanowires.
\end{abstract}


\maketitle


There is growing interest in superconducting nanowire single-photon detectors (SNSPDs) fueled by their combination of good detection efficiency, low dark-count rate, very short recovery time and exceptionally small jitter. SNSPD compare well with other competing technologies for the detection of visible and near-infrared photons \cite{Hadfield09} and have already been used in a wide variety of applications from quantum key distribution to time-of-flight depth ranging \cite{Natarajan12}. Variations of these detectors have also been used to detect higher energy particles, such as high kinetic-energy molecules in mass spectrometry \cite{Suzuki08,Ohkubo12} or x-ray photons with keV-energies \cite{Inderbitzin12,Inderbitzin13}.

The active element of these detectors consists of a typically square meander of a superconducting NbN film of a few nanometer thickness \cite{Goltsman01}. Recently alternative superconducting materials have been suggested, such as NbTiN \cite{Dorenbos08}, TaN \cite{Engel12}, or WSi \cite{Baek11}, which may be better suited than NbN for certain applications. The detectors are biased with a constant direct current \Ibias\ which usually equals about 90\% to 95\% of the experimental critical current $I_c$. The absorption of a photon of sufficient energy, in combination with a suitably chosen bias current, can trigger a normal conducting cross-section, the subsequent growth of which is determined by the electro-thermal properties of the detector \cite{Gurevich87}. Estimates\cite{Semenov05a,Kerman09} and simulations\cite{Marsili11} made with realistic material and device parameters for SNSPD show that this normal conducting domain initially grows very fast with a correspondingly large resistance. As a consequence part of the bias current is redirected into the readout line, which effectively acts as a $50~\Omega$ parallel impedance, or an additionally installed parallel ohmic resistance, thus reducing the Joule-heating. Depending on the details of this electro-thermal feedback \cite{Kerman09}, the detector can be operated in  the desired self-recovering mode or it latches into a resistive state, when the normal-conducting domain is stabilized by self-heating. This electro-thermal feedback imposes limits on the minimum achievable recovery time and the maximum count rate.

Many further aspects of SNSPD relevant for applications are also well understood. State-of-the-art SNSPD consist of a homogeneous superconducting film and a uniform meander without constrictions. Such devices exhibit a nearly constant detection efficiency for a certain range in parameter space of bias current and photon energy. In this ``plateau region'' nearly every absorbed photon triggers the formation of a normal conducting domain \cite{Kerman07} leading to an intrinsic detection efficiency ($IDE$) approaching 100\% \cite{Hofherr10,Lusche13}. The device detection efficiency ($DDE$) is the product of photon absorptance ($ABS$) of the meander and IDE: $DDE = ABS \times IDE$. $ABS$ itself depends on the absorptance of the thin superconducting film and geometric effects, such as the meander filling factor or a polarization dependent absorptance \cite{Dorenbos08a} and is limited to $\lesssim 20\%$ for a bare meander. Higher $DDE>50\%$ can be achieved by incorporating the detector into an optical cavity \cite{Rosfjord06}, and optimizations of layer thicknesses and the separation of the meander lines allow for high detection efficiencies even at relatively low filling factors \cite{Akhlaghi13,Yamashita13}. The highest reported system detection efficiency of up to $93\%$ reported to date, including coupling losses and photon absorption in the optical fibre, was achieved for an optimized WSi SNSPD \cite{Marsili13}. For a given bias current there is a minimum threshold energy for photons to be detected with the maximum efficiency. Photons with wavelengths $\lambda$ larger than the corresponding cut-off wavelength $\lambda_c$ can only be detected with a rapidly decreasing probability. Narrower meander lines \cite{Marsili11a} or superconducting films with a lower $T_c$ \cite{Baek11,Dorenbos11,Engel12} result in an increase of $\lambda_c$ and higher detection efficiencies at long photon wavelengths.

Detector noise or so called dark counts increase approximately exponentially on approaching the experimental critical current. Experimental \cite{Bartolf10} and theoretical investigations \cite{Bulaevskii11} have favored magnetic vortices crossing the superconducting strips as the dominant mechanism leading to intrinsic dark-count events. Except near the ends of the straight sections of the superconducting meander the current density in the undisturbed equilibrium situation is homogeneous due to the fact that the strip width $w<<\Lambda=2\lGL^2/d$, where $\Lambda$ is the effective 2D magnetic penetration depth, $\lGL\gg d$ is the corresponding Ginzburg-Landau (GL) magnetic penetration depth in the bulk material and $d$ the film thickness.
Due to a current-crowding effect the current density in the $180^\circ$ turnarounds of the meander structure is no longer homogeneous\cite{Clem11} and dark-count events most likely originate near these turnarounds \cite{Engel12a}, but may be reduced by a more sophisticated meander design \cite{Akhlaghi12a}.

Despite this remarkable progress in understanding and optimization of SNSPD some open questions remain, particularly in connection with the mechanism that is responsible for triggering the initial resistive cross-section. The first model \cite{Semenov01} describing the detection mechanism in SNSPD assumed the formation of a normal-conducting hot-spot that diverts the applied bias current into the still superconducting side-walks. The current density in these side-walks will eventually increase beyond the critical current density, thereby leading to the initial normal-conducting cross-section. This model, which we will call hard-core model, already captures some important characteristics of SNSPD, such as the existence of a bias-current dependent minimum photon energy, and because it gives a very vivid description, it has been widely used. However, it fails to describe certain observations, \emph{e.g.}\ the temperature dependence of $\lambda_c$ \cite{Engel13}, and often leads to inconsistencies. As an example we cite data from a recent analysis of cut-off wavelengths within the hard-core model \cite{Maingault10}. The corresponding analysis results in a hot-spot diameter of about $12$~nm for a photon energy of $1.24$~eV. Using the device parameters given in Ref.~\onlinecite{Maingault10} and typical superconductivity parameters for NbN one can estimate the superconducting condensation energy of the corresponding volume to be of the order of $0.1$~eV. In the same paper, the energy conversion efficiency was estimated to $\zeta\approx0.5\%$, which means that it would require a photon energy at least one order of magnitude larger than used in the experiment to drive this volume into the normal conducting state.

One inherent shortcoming of the hard-core model originates from neglecting excess quasi-particles (QP) outside the normal-conducting core in the still superconducting side-walks. An alternative detection model has been suggested several years ago \cite{Engel05,Semenov05a} in which the reduction of the depairing critical current in a cross-section of a superconducting strip due to excess QP is taken into account. This QP-model explicitly does not require a normal conducting region to form before the critical current has dropped below the applied bias current. Very recently it has been suggested that the relevant current scale is not the depairing critical current but instead the critical current for vortex crossings \cite{Bulaevskii12}, although there is some controversy about the correct theoretical treatment of a vortex very near the strip edge\cite{Gurevich12}. Also, more advanced theoretical models based on the time-dependent GL theory are being developed \cite{Zotova12} with the aim to gain a better understanding of the dynamics of the detection mechanism. A full numerical simulation based on the time-dependent GL theory coupled with heat diffusion and the Maxwell equations \cite{Ota13} resulted in threshold energies required to trigger a normal conducting domain at least one order of magnitude larger than experimentally observed.

In this paper we present the development of simple model of the QP multiplication and diffusion process that is detailed enough to allow a comparison with experimental results. In Sec.~\ref{Sec.model} we develop the mathematical model that allows us to numerically determine detection criteria for direct photon detection within the hard-core, QP, and vortex model. The corresponding detection criteria will also be specified. In Sec.~\ref{Sec.parameters} the material and geometric parameters used in our simulations are defined and we perform some consistency tests to validate our results. This will be followed by the presentation of our numerical results and a comparison with experimental data whenever this is possible.

\section{Development of the physical model\label{Sec.model}}

\subsection{Modelling the quasi-particle diffusion}

We will restrict ourselves to a discussion of the detection of mainly visible and near-infrared photons with energies $h\nu\sim1$~eV. These energies are much larger than the superconducting gap $\Delta\sim 1$~meV of typical SNSPD. Absorption of such a photon results in a large number of excitations in the form of QPs and phonons. Details of this QP multiplication process have been already described in an early publication on SNSPD \cite{Semenov01} and references therein. In this paper we are not interested in the detailed time-evolution of the number of excited QPs, instead we will resort to a simple analytical approximation. We make the assumption that the time scale of electron-electron interactions is much faster than both the electron-phonon and phonon-phonon time scales\cite{Semenov01}. This means that the electronic system will have thermalized to a local, (near-)equilibrium state long before the phonon system thermalizes.
We will also assume that the increase in the local concentration of QPs equals the local reduction in the concentration of superconducting electrons, in other words, that the electronic system is always in a local, near-equilibrium state, and we neglect the background of thermally excited QPs\footnote{Calculations were done at $T/T_c=0.05$ resulting in a very low concentration of thermal QPs}.

Previous models \cite{Semenov01,Semenov05a} have made very similar assumptions and furthermore assumed the QP multiplication and QP diffusion to be independent processes. Instead, we will be considering the highly excited electron after photon absorption, which itself diffuses within the superconducting film, to be the source of QPs as it continuously loses energy, thereby breaking up Cooper-pairs. Assuming that the thermalization process does not influence the diffusion of the excited electron, the probability density $C_e(\vec{r},t)$ to find it at position $\vec{r}$ at time $t$ after the absorption follows the diffusion equation
\begin{equation}
\frac{\partial C_e(\vec{r},t)}{\partial t} = D_e\nabla^2 C_e(\vec{r},t),
\label{Eq.PDE_e}
\end{equation}
with $\nabla^2$ the Laplace-operator and $D_e$ the diffusion coefficient of normal electrons. We set the diffusion coefficient to be constant. In reality it may be a function of the excitation energy of the electron, thus depending on $t$ and the photon energy. Already after $t\lesssim 1$~ps the diffusion length $\sim\sqrt{D_{e}t}$ becomes larger than the typical superconducting film thickness $d\approx5$~nm. It is therefore justified to treat this problem in two dimensions for longer time scales. During the diffusion process the electron thermalizes by losing energy in inelastic scattering events. We simplify this process assuming an exponential decay of the excitation energy with a constant time scale $\tau_{qp}$: $E_e=h\nu\exp\left(-t/\tau_{qp}\right)$. A certain fraction of these scattering events results in the generation of two QPs with an energy $\Delta$ and is proportional to the probability density $C_e(\vec{r},t)$.

The excess QPs themselves undergo diffusion in the superconducting film before they recombine to form Cooper-pairs on a time scale $\tau_r\gg\tau_{qp}$. The concentration of excess QPs $C_{qp}(\vec{r},t)$ can then be described by
\begin{widetext}
\begin{equation}
  \frac{\partial C_{qp}(\vec{r},t)}{\partial t} = D_{qp}\nabla^2 C_{qp}(\vec{r},t) - \frac{C_{qp}(\vec{r},t)}{\tau_r} + \frac{\zeta h\nu}{\Delta\tau_{qp}}\exp\left(-\frac{t}{\tau_{qp}}\right)C_e(\vec{r},t),
\label{Eq.PDE_QP}
\end{equation}
\end{widetext}
with $D_{qp}\neq D_e$ the QP diffusion coefficient and $0<\zeta\leq1$ the conversion efficiency, which has to be determined experimentally. The last term in Eq.~\eqref{Eq.PDE_QP} is the source term describing the QP-multiplication process. The recombination process is described by $\frac{C_{qp}(\vec{r},t)}{\tau_r}$, which we include in a linear approximation \footnote{It has to be expected that the recombination is proportional to $C_{qp}(\vec{r},t)^2$, since two QPs are required. Because $\tau_r\gg\tau_{qp}$ recombination has a minor influence on the detection process which happens on a time-scale $\sim\tau_{qp}$. Therefore we use the linear approximation resulting in a set of linear differential equations.}. Eqs.~\eqref{Eq.PDE_e} and \eqref{Eq.PDE_QP} form a set of coupled differential equations describing the evolution of the statistically averaged density of excess QPs.

\begin{figure}
 \includegraphics[width=\columnwidth,totalheight=200mm,keepaspectratio]{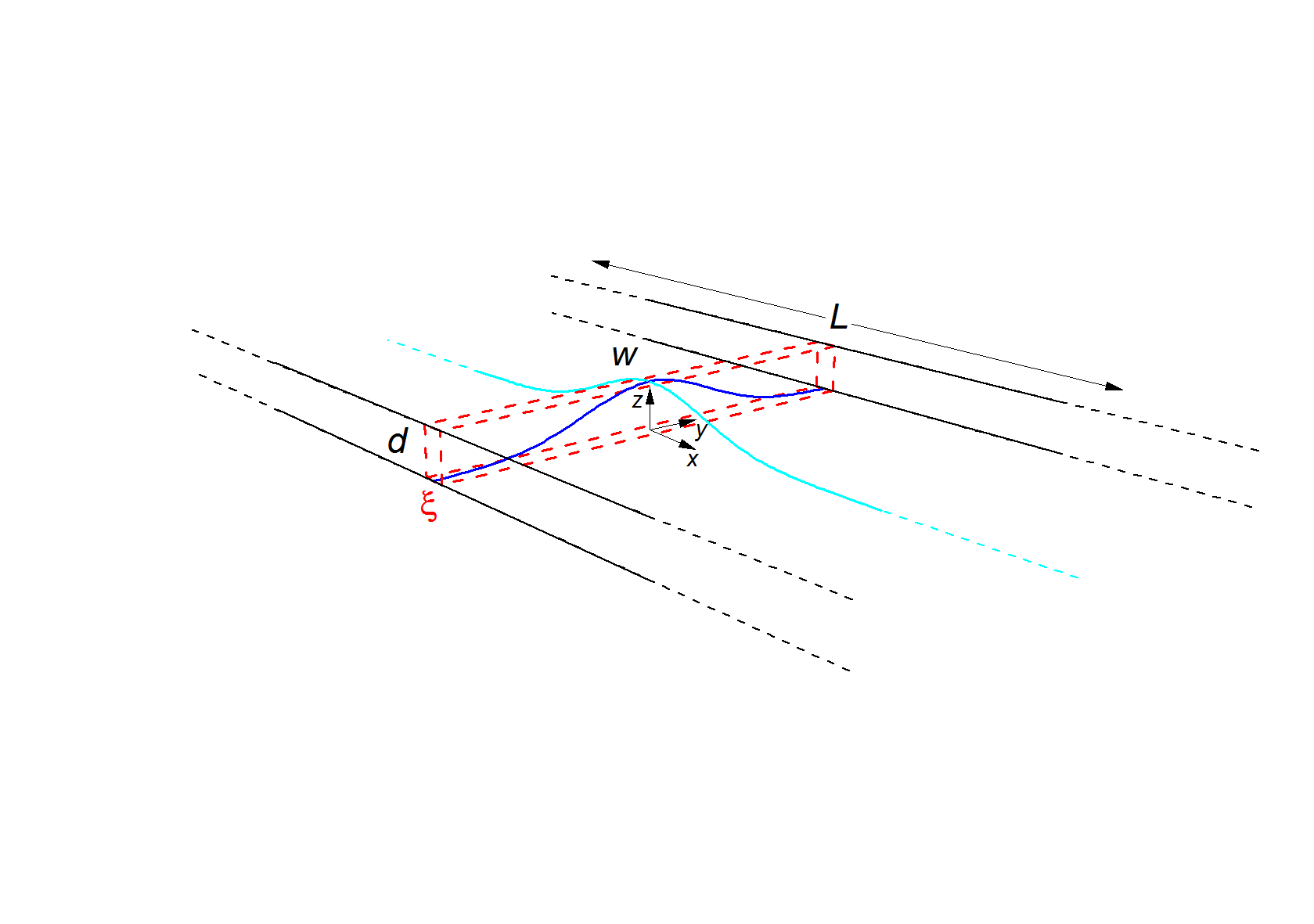}%
 \caption{Schematic drawing of meander section with the photon absorption site at its center, not to scale. Indicated is the $\xi$-slab that defines the minimum volume that has to switch into the normal conducting state. Also plotted are Gaussian-profiles (dark and light blue) of the QP concentration according to Eq.~\eqref{Eq.AnalyticalSolution} at an arbitrary time $t$.\label{Fig.Schematic}}
\end{figure}

For the case of an infinitely large 2D-film and making the assumption $D_{qp}= D_e=D$, Eqs.~\eqref{Eq.PDE_e} and \eqref{Eq.PDE_QP} have an analytical solution (see Appendix \ref{App.Analytic}),
\begin{widetext}
\begin{align}
C_e(r,t) &= \frac{1}{4\pi Dt}\exp\left(-\frac{r^2}{4Dt}\right),\label{Eq.2D_Diffusion}\\
C_{qp}(r,t) &= \frac{\zeta h\nu}{\Delta}\frac{\tau_r}{\tau_r-\tau_{qp}}\left[\exp\left(-\frac{t}{\tau_r}\right)-\exp\left(-\frac{t}{\tau_{qp}}\right)\right]\frac{1}{4\pi Dt}\exp\left(-\frac{r^2}{4Dt}\right),
\label{Eq.AnalyticalSolution}
\end{align}
with $r$ being the distance from the absorption site of the photon. Integration of Eq.~\eqref{Eq.AnalyticalSolution} over the complete film gives the total number of excess QPs,
\begin{equation}
N_{qp}(t) = \int_\infty C_{qp}(\vec{r},t)\mathrm{d}V = \frac{\zeta h\nu}{\Delta}\frac{\tau_r}{\tau_r-\tau_{qp}}\left[\exp\left(-\frac{t}{\tau_r}\right)-\exp\left(-\frac{t}{\tau_{qp}}\right)\right].
\label{Eq.totalQPnumber}
\end{equation}
\end{widetext}
This last equation also holds for the more realistic case of narrow superconducting strips and $D_e \neq D_{qp}$ as long as $\tau_{qp}$ is independent of $C_{qp}$. However, the concentration $C_{qp}(\vec{r},t)$ itself has to be calculated numerically in this more general situation. In the following we will solve this set of differential equations \eqref{Eq.PDE_e} and \eqref{Eq.PDE_QP} for a rectangle of width $w$ and length $L\gg w$, see Fig.~\ref{Fig.Schematic}. For the side-walls we use Neumann boundary conditions $\partial C(\vec{r},t)/\partial y = 0$, \emph{i.e.}\ no loss of QP through the side-walls, and perfectly absorbing walls at the beginning and end of the strip. We assume the photon to be absorbed in the center of the strip. The numerical solution to Eqs.~\eqref{Eq.PDE_e} and \eqref{Eq.PDE_QP} is found using the finite element method (FEM) and Matlab$^\mathrm{TM}$ software. The mesh on which the solutions are calculated was created with a high density of nodes around the absorption site. For the subsequent analysis of the trigger models the results are transformed onto a cartesian grid with a resolution of $0.5$~nm in $x$- and $y$-direction. A finer grid of $0.2$~nm resolution was used in a central region $\pm5\xi(0)$ around the absorption site in $x$-direction and over the full width of the strip. The parameters $\xi$, \lGL, $\Delta$ and $D_{qp}\neq D_e$ are in general assumed to be temperature dependent (see Appendix \ref{App.T-dependence} for details).

\subsection{Detection criteria}

Based on the computed $C_{qp}(\vec{r},t)$, different detection criteria can be formulated and compared with each other. In the following we will consider three models that explain the formation of the initial normal conducting cross-section. The first and original model describing photon detection in SNSPD \cite{Semenov01} assumes QP-concentrations high enough to completely suppress superconductivity in the vicinity of the absorption site. For this hard-core model we use a condition similar to Ref.\ \onlinecite{Semenov01}, namely we define the extension of the normal conducting core of the hot-spot by $C_{qp}(\vec{r},t)\geq n_{se}^{2D}=n_{se}d$, with $n_{se}$ the equilibrium density of superconducting electrons\footnote{The density of superconducting electrons $n_{se}$ has been estimated from the London penetration depth $\lambda_L^2=m_e/(\mu_0n_{se}e^2)$.} in the film being twice the Cooper-pair density, and we require this normal-conducting area to have an extension of at least the coherence length $\xi$ in the direction of the applied current. The current-density inside the normal conducting area is assumed to be zero, thus leading to an enhanced current-density in the side-walks as required by the continuity equation. If the current density in the side-walks exceeds the depairing critical current \Idep, the whole cross-section becomes normal conducting leading to the detection of the absorbed photon. The minimum transversal extension $2\Rhc$ of the normal conducting core for photon detection, the strip width $w$ and the reduced bias current $\Ibias/\Idep$ are then related to each other defining the detection criterion
\begin{equation}
2\Rhc \geq w\left(1-\frac{\Ibias}{\Idep}\right).
\label{Eq.HardCoreCriterium}
\end{equation}
If one assumes the conversion efficiency $\zeta\lesssim 1$, near-infrared photons with $h\nu\lesssim1$~eV would produce normal conducting areas large enough to be roughly consistent with experimental results. However, more plausible values \cite{Ilin00,Semenov05a,Engel12} for the conversion efficiency are on the order of $0.1$, in which case correspondingly higher photon energies are required.

The most important conceptual short-coming of the hard-core model is the restriction of the QPs to be strictly confined to the potentially present normal conducting volume. For very high photon \cite{Inderbitzin12} or particle energies \cite{Suzuki11} the contribution of excess QPs outside the normal conducting volume is probably negligible, but they may become significant or even dominating for smaller excitation energies. This was first realized in the QP-model \cite{Semenov05a} which does not require a normal conducting volume. Instead, an excess number of QPs results in a reduction of the effective critical current. Under the assumption that the number of excess QPs equals the reduction in the number of superconducting electrons, the effect on the critical current can be easily calculated applying the continuity equation for the applied bias current and the phase coherence of the superconducting electrons. The minimum volume that must reach the normal state to trigger a photon count has to have a length in the direction of the applied current of at least the coherence length and span the complete cross-section, the ``$\xi$-slab'', see also Fig.~\ref{Fig.Schematic}. As a condition for the nucleation of such a normal conducting cross-section one obtains \cite{Semenov05a}
\begin{equation}
\frac{N_{qp}^\mathrm{slab}(t)}{N_{se}} \geq 1-\frac{\Ibias}{\Idep},
\label{Eq.CriticalCurrentReduction}
\end{equation}
where $N_{se}=n_{se}wd\xi$ is the equilibrium number of superconducting electrons in the $\xi$-slab. The number of excess QPs in the $\xi$-slab is computed from
\begin{equation}
N_{qp}^\mathrm{slab}(t)=\int_\mathrm{\xi-slab}C_{qp}(\vec{r},t)\mathrm{d}V.
\label{}
\end{equation}

As a third possibility that could lead to the formation of a normal conducting domain, we consider the photon-assisted crossing of a vortex \cite{Bulaevskii12}. In the case of a homogeneous current density $\jbias=\Ibias/w=\mathrm{const.}$ across the strip, it is straightforward to calculate the energy barrier prohibiting the entry and subsequent crossing of vortices \cite{Bulaevskii11,Clem11}. Contrary to the original publication\cite{Bulaevskii12}, where the authors assume a uniformly reduced order parameter, we consider an inhomogeneous current density due to the expanding cloud of QPs after photon absorption. We assume the current redistribution to be instantaneous, justified by an estimate of the GL relaxation time $\tau_\mathrm{GL}\lesssim1$~ps\cite{Kopnin01}. We can then calculate the local, time-dependent current density proportional to the local density of superconducting electrons, $n_{se}^{2D}-C(\vec{r},t)$, and require current continuity\footnote{If a normal conducting core forms, we set the local density of superconducting electrons equal zero inside the core.} and $\operatorname{div} \vec{j}=0$. This leads to an enhancement of the current density near the strip edges and a corresponding decrease of the energy barrier for vortex entry very similar to the situation near the meander turn-arounds \cite{Clem11}. We calculate the forces on a vortex as a function of its position inside the strip taking into account the inhomogeneous current distribution as well as the increase of the effective penetration depth $\Lambda$ due to the reduced density of superconducting electrons and obtain the vortex potential by numerical integration (see Appendix \ref{App.VortexPotential}). For a given reduced bias current $\Ibias/I_{c,v}$ the energy barrier vanishes for a minimum photon energy, giving us the cut-off wavelength $\lambda_c$ in this vortex-model. It is important to note that the current scale in the vortex model is $I_{c,v}<\Idep$, \emph{i.e.}\ the current for which the energy barrier is reduced to zero \cite{Bulaevskii11}.

The model considered by Zotova and Vodolazov\cite{Zotova12} is closely related to this vortex model. These authors assume the formation of a normal-conducting core similar to the hard-core model, but take into account a non-homogeneous current distribution around the normal-conducting area due to a current-crowding effect\cite{Clem11}. In such a situation the highest current density is expected at a point very close to this normal-conducting area inside the strip, favoring the creation of a vortex-antivortex pair. In case of a continuous variation of the Cooper-pair density, we expect this current-crowding effect to be much less pronounced. Unfortunately, there is, to our knowledge, no simple method available to calculate the exact current distribution in this more complicated situation. The method we used to calculate the current densities does not lead to an enhanced current density close to the photon absorption site, and we therefore do not consider the creation of vortex-antivortex pairs in our analysis.

Before discussing our results for the three models, we state that our numerical model contains a number of simplifications in addition to our general assumptions discussed at the beginning of this section, the most important ones we would like to mention here. We assume the superconducting gap $\Delta$ to be independent of the density of excess QP, but it is known that the presence of QPs leads to a certain reduction of $\Delta$ \cite{Gilabert90}, the magnitude of which depends on the details of the thermalization process. This leads to an underestimation of the effects that an absorbed photon causes in the superconducting strip, and thus to an overestimation of the minimum photon energy required to trigger a detection event in all three detection models, but in general to a different degree. A presumably smaller error is introduced by neglecting the reduction of $\Delta$ due to $\Ibias$. When comparing our simulation results with experimental measurements, we can partially correct these effects by assuming a higher effective conversion efficiency $\zeta$, which is an adjustable parameter with no well established theoretical estimates. Further significant simplifications are made for the QP multiplication process, for example, the assumption of a constant time-scale $\tau_{qp}$ for the QP-multiplication. Our assumption of a nearly unchanged concentration of Cooper-pairs is also not strictly fulfilled during the early stages of the QP multiplication and diffusion process near the absorption site, since our simulations indicate high concentrations of QPs for $t\lesssim1$~ps and correspondingly small Cooper-pairs concentrations even for photon wavelengths much longer than $\lambda_c$.

\section{Simulation parameters and consistency checks}\label{Sec.parameters}

In Table \ref{Tab.Material} we summarize typical values of material parameters for high-quality films of TaN \cite{Engel12} and NbN \cite{Bartolf10} as we used them in the calculations. The parameters $\Delta$, $\xi$, \lGL and $D_{qp}$ are assumed to be temperature dependent (see Appendix \ref{App.T-dependence} for details). The temperature dependent value of $D_{qp}$ is calculated from $D_e$ of the normal-conducting electrons at $T_c$ as detailed in Appendix \ref{App.T-dependence}. For the current report, we have set $T/T_c=0.05$. The general temperature-dependent behavior of SNSPD is the subject of ongoing investigations and will be presented in a future publication. The time constant $\tau_r$ has been chosen as an average value for the whole $T$-range \cite{Semenov97}, and as it turns out, our results are not sensitive to the choice of $\tau_r$. The time constant $\tau_{qp}$ is related to the thermalization time $\tau_{th}$, which has been measured for NbN to be $\approx7$~ps \cite{Ilin98}. With the chosen time constants the total maximum number of excess QPs is reached at $t=\tau_{th}\approx10.3$~ps, as calculated using Eq.~\eqref{Eq.totalQPnumber}, and reaches $\gtrsim98$\% of this maximum number of QPs at $t=7$~ps. Table \ref{Tab.Geometry} lists the geometrical parameters of the simulated superconducting strip.

\begin{table*}
\caption{\label{Tab.Material} Summary of material parameters for TaN and NbN (for $T=0.05$) that are important for the comparison with analytical approximations and are entering the simulation.}
\begin{tabular}{lccccccccc}
 & $\Delta$ (meV) & $\xi$ (nm) & \lGL\ (nm) & $D_e$ (nm$^2$ ps$^{-1}$) & $D_{qp}$ (nm$^2$ ps$^{-1}$) & $\zeta$ & $\tau_{qp}$ (ps) & $\tau_r$ (ps) & $N_0$ (nm$^{-3}$eV$^{-1}$)\\
 \hline
 TaN & 1.3 & 5.3 & 520 & 60 & 8.2 & 0.25 & 1.6 & 1000 & 48 \\
 NbN & 2.3 & 4.3 & 430 & 52 & 7.1 & 0.25 & 1.6 & 1000 & 51
\end{tabular}
\end{table*}

\begin{table}
\caption{\label{Tab.Geometry} Geometric parameters used for the simulations. The FEM solutions have been transformed from the simulation mesh to a cartesian grid for subsequent analyses. Over the whole strip a rough grid was applied, and in an area $\pm5\xi(0)$ around the absorption site and spanning the width of the strip a finer grid was used.}
\begin{tabular}{lr}
Length $L$ & $1\ \mu$m \\
Width $w$ & $100$ nm \\
Thickness $d$ & $5$ nm \\
rough grid $\Delta x,\Delta y$ & $0.5$ nm \\
fine grid $\Delta x,\Delta y$ & $0.2$ nm
\end{tabular}
\end{table}

\begin{figure}
 \includegraphics[width=\columnwidth,totalheight=200mm,keepaspectratio]{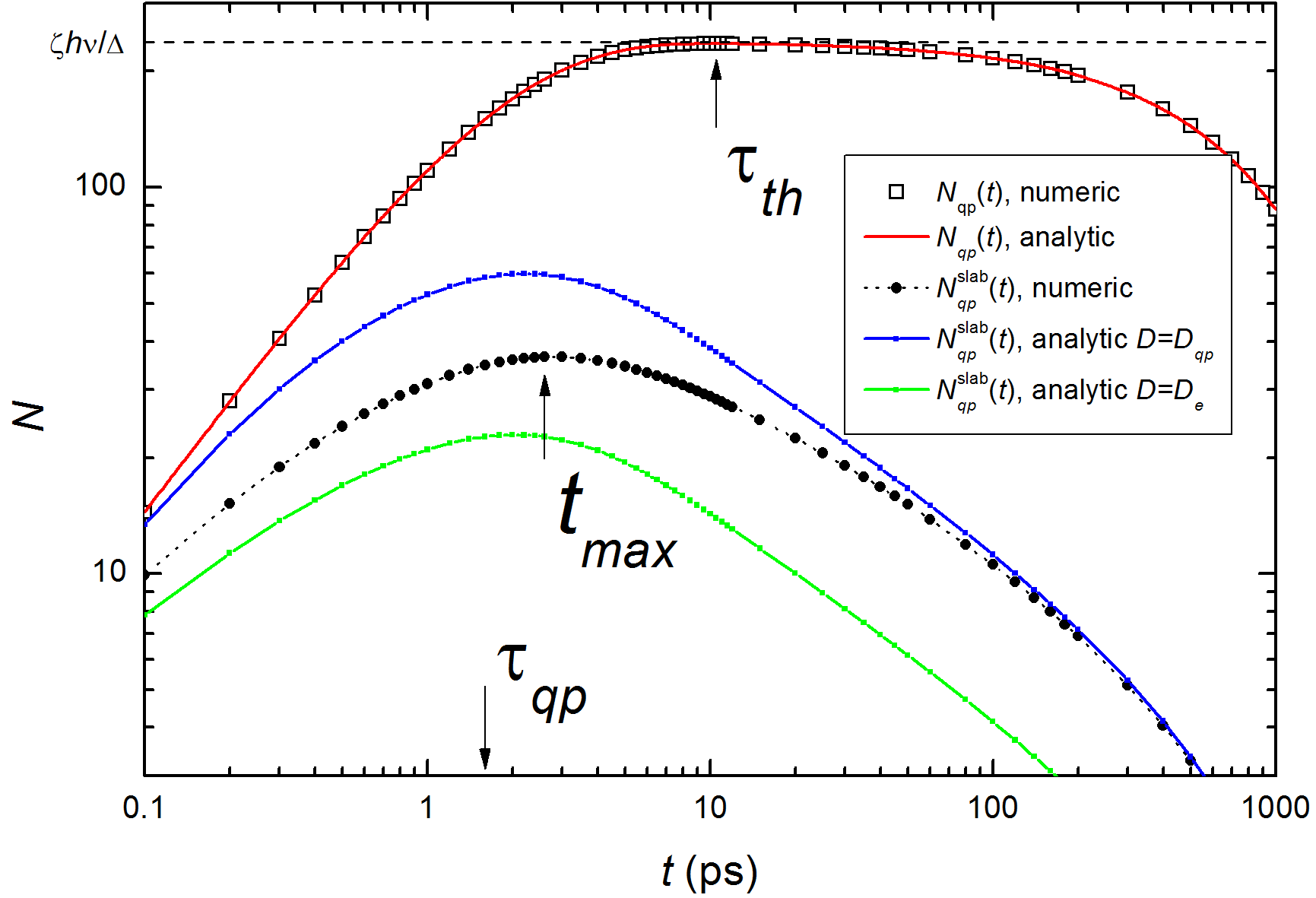}%
 \caption{Double-logarithmic plot of the number of QPs in the complete superconducting strip $N_{qp}(t)$ (black squares and solid red line) and within the $\xi$-slab $N_{qp}^\mathrm{slab}(t)$ (black, blue and green) as functions of time, calculated for a photon wavelength with $\lambda=1000$~nm absorbed in a TaN-film. Numeric results are compared to analytical approximations Eqs.~\eqref{Eq.totalQPnumber} and \eqref{Eq.N_xislab}. Arrows indicate the times of maximum total number of QPs, $\tau_{th}$, maximum number of QPs in the $\xi$-slab, $t_{max}$, and QP multiplication time scale, $\tau_{qp}$. Also indicated is the asymptotic maximum number of QPs $\zeta h\nu/\Delta$ for the theoretical case of no recombination of QPs into Cooper-pairs (dashed horizontal black line). \label{Fig.Nvont}}
\end{figure}

We verified the validity of our numerical calculations by comparing the results to those using analytical expressions for an infinite film based on Eq.~\eqref{Eq.AnalyticalSolution}. Fig.~\ref{Fig.Nvont} shows the temporal evolution of the number of QPs in the complete strip as well as in the $\xi$-slab on a double logarithmic scale. Calculations were done for a photon with wavelength $\lambda=1000$~nm absorbed in a TaN-film. The solid red line is the calculated number of QPs $N_{qp}(t)$ according to Eq.~\eqref{Eq.totalQPnumber} and the black squares represent the same quantity obtained by numeric integration of $C_{qp}(\vec{r},t)$ over the complete strip. The numeric results agree very well with the analytical expression (better than $1$\% for all $t<=1$~ns) and the thermalization time $\tau_{th}=10.5$~ps also agrees with the analytic result of $10.3$~ps within the temporal resolution of the simulation ($\pm0.5$~ps for $3~\text{ps}\leq t\leq12$~ps).

In the same figure we also plotted the numerically calculated number of QPs in the $\xi$-slab $N_{qp}^\mathrm{slab}(t)$ (black circles). It shows a pronounced maximum at $t_{max}\approx2.6$~ps after absorption of the photon, significantly before the total number of excess QPs have reached their maximum at $\tau_{th}$. According to Eq.~\eqref{Eq.CriticalCurrentReduction} $t_{max}$ corresponds to the moment when a certain minimum bias current can still trigger the formation of the initial normal conducting cross-section in the QP-model. An analytic solution can be found using Eq.~\eqref{Eq.AnalyticalSolution} and approximating the integration in the $x$-direction by $\tanh(\xi/\sqrt{4\pi Dt})$. The approximate number of QPs in the $\xi$-slab then becomes
\begin{equation}
N_{qp}^\mathrm{slab}(t) \approx N_{qp}(t)\tanh\left(\frac{\xi}{\sqrt{4\pi Dt}}\right).
\label{Eq.N_xislab}
\end{equation}
In Fig.~\ref{Fig.Nvont} $N_{qp}^\mathrm{slab}(t)$ according to Eq.~\eqref{Eq.N_xislab} is plotted for $D=D_{qp}$ (blue symbols) and $D=D_e$ (green symbols), respectively. For $t\gtrsim100$~ps the analytic solution with $D=D_{qp}$ asymptotically approaches the numerical solution. For smaller $t$ neither solution gives an adequate description of the numerical results, thus demonstrating the necessity of numerical calculations. The physical reason is the distinction between the diffusion coefficients for normal electrons and QPs, respectively. At the beginning of the multiplication process QPs are generated proportional to $C_e(r,t)$ with $D_e>D_{qp}$. However, the QPs immediately start to diffuse with $D_{qp}$. The effective diffusion coefficient becomes $t$-dependent and approaches $D_{qp}$ once the generation of additional QPs has stopped for $t\gg\tau_{qp}$.

\begin{figure}
 \includegraphics[width=\columnwidth,totalheight=200mm,keepaspectratio]{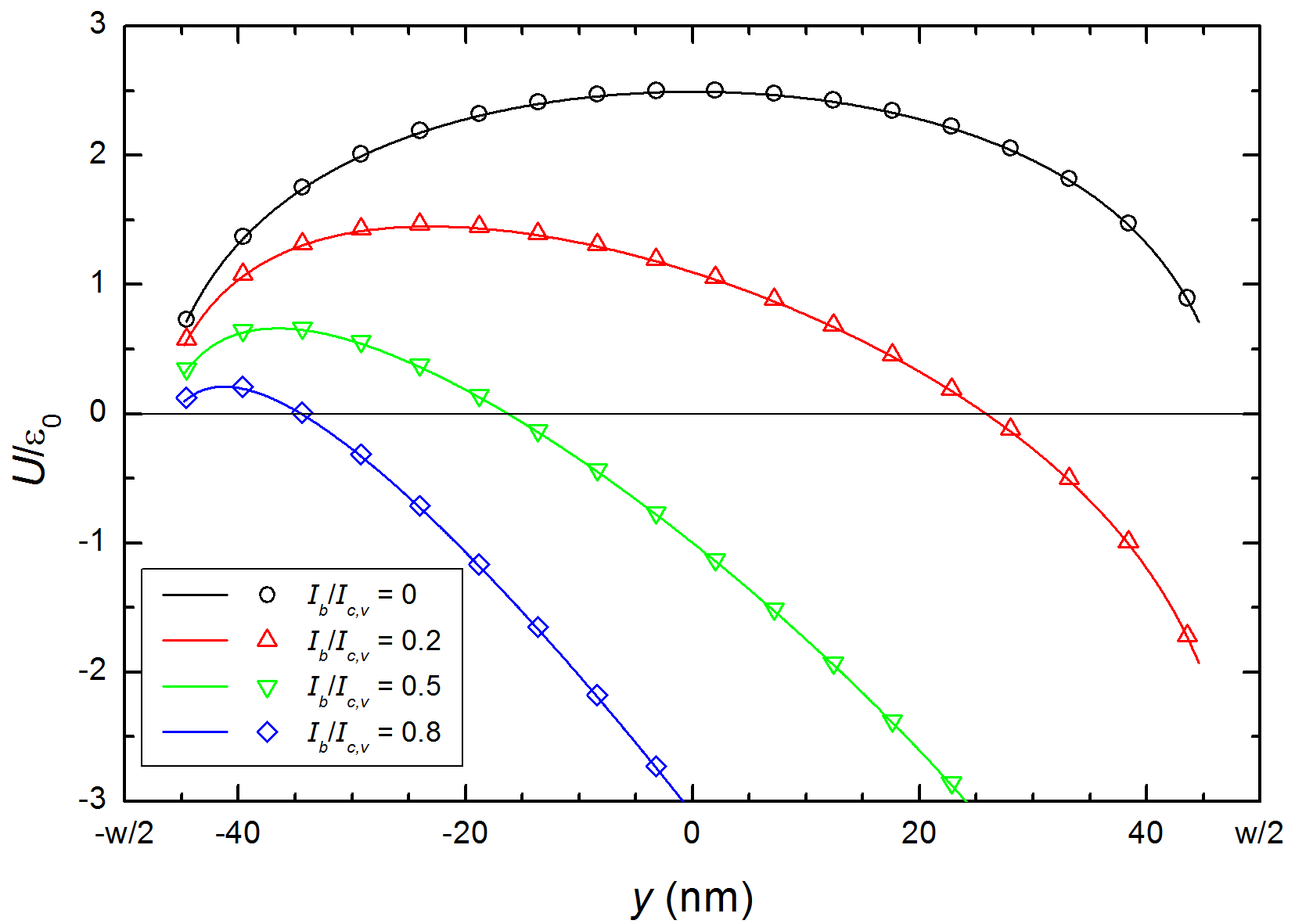}%
 \caption{Potential energies of single vortices, as functions of the position across the strip, calculated numerically (open symbols) and according to Eq.~\eqref{Eq.VortexPotential} (solid lines) for different bias currents as indicated. Calculations were done for homogeneous current densities and $T/T_c=0.05$.\label{Fig.VortexPotEquilibrium}}
\end{figure}

Finally we compare our numerically obtained potential energies of single vortices as a function of position $y$ across the strip with the analytical expression for the case of a homogeneous current density \cite{Bulaevskii12}
\begin{equation}
U(y,I,T)/\varepsilon_0 = \ln\left[\frac{2w}{\pi\xi(T)}\cos\left(\frac{\pi y}{w}\right)\right] - \frac{I}{I_{c,v}}\frac{2(y+\frac{w}{2})}{\exp(1)\xi(T)},
\label{Eq.VortexPotential}
\end{equation}
where $\varepsilon_0=\Phi_0^2/(2\pi\mu_0\Lambda)$ is the characteristic vortex energy, $\Phi_0=h/2e$ the magnetic flux quantum and $\mu_0$ the permeability of free space. Eq.~\eqref{Eq.VortexPotential} has been derived for the condition $d\ll w\ll\Lambda$, which is typically fulfilled for SNSPD. We follow here Ref.~\onlinecite{Bulaevskii12} and set the first term on the right hand side in Eq.~\ref{Eq.VortexPotential} to zero at $y=(\xi(T)-w)/2$ and set $U(y,I,T)=0$ for $y<\xi-w/2(T)$ and $y>w/2-\xi(T)$. We plot in Fig.~\ref{Fig.VortexPotEquilibrium} the analytical and numerical results for $T/T_c=0.05$ and four values of $\Ibias/I_{c,v}$ as indicated. Again, numerically and analytically obtained values agree with each other to within a few percent.

\section{Simulation results and comparison with experimental data}

\subsection{Temporal evolution of QP density, current distribution and vortex edge-barrier}

\begin{figure}
 \includegraphics[width=\columnwidth,totalheight=0.8\textheight,keepaspectratio]{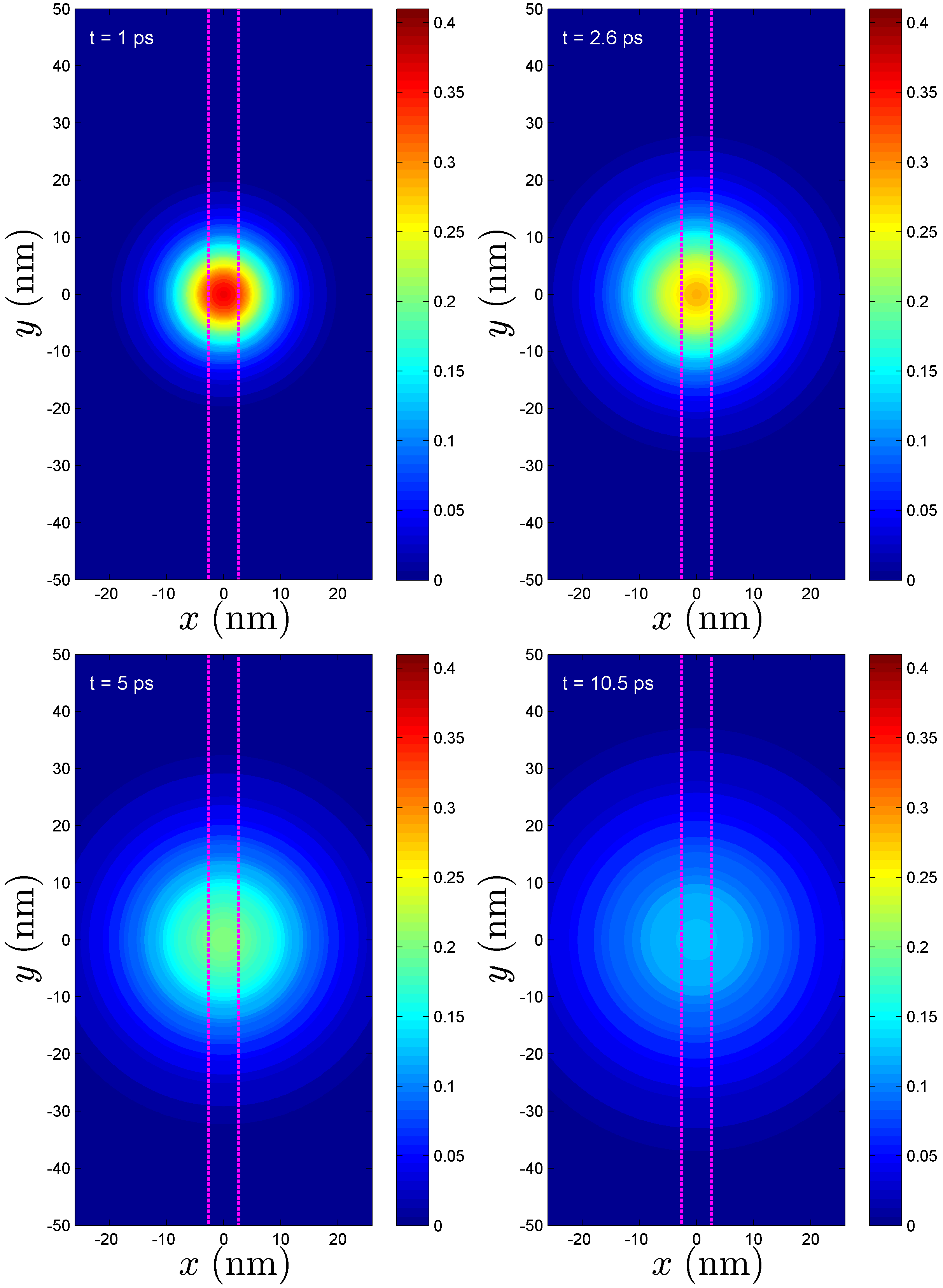}%
 \caption{QP density for different times after absorption of a $1000$~nm photon at $T/T_c=0.05$ in the center of the superconducting TaN strip. In panel (b) the number of QPs reaches its maximum in the $\xi$-slab ($t=t_{max}=2.6$~ps), which is indicated by the vertical dashed lines. Panel (d) shows the situation when the maximum number of QPs for the complete strip is reached at $t\approx10.5$~ps. However, at this point diffusion has led to a significant reduction of QP in the $\xi$-slab. Panels (a) and (c) depict the situation at times before and after $t_{max}$.\label{Fig.QPdensity}}
\end{figure}

The primary goal of our simulations is to reveal the temporal evolution of the excess QP density in the superconducting strip. In Fig.~\ref{Fig.QPdensity} we show typical results for the case when superconductivity is not completely suppressed in the hot-spot core. These calculations were done with material parameters for TaN at a reduced temperature $T/T_c=0.05$ and an incident photon with $\lambda=1000$~nm absorbed in the center of the strip. At the very early stages after photon absorption ($t=1$~ps, panel (a) in Fig.~\ref{Fig.QPdensity}) the QPs are highly concentrated near the absorption site, leading to a significant suppression of superconductivity in a very small volume. Already at $t=t_{max}=2.6$~ps the maximum number of QPs in the $\xi$-slab is reached (compare to Fig.~\ref{Fig.Nvont}), see the situation shown in panel (b) of Fig.~\ref{Fig.QPdensity}. Despite the relatively low diffusion coefficient $D_{qp}$ at this low temperature, a significant number of QPs has diffused out of the $\xi$-slab, and although the total number of excess QPs continues to increase until $t=t_{th}\approx10.5$~ps, the concentration of QPs in the $\xi$-slab drops more quickly, resulting in a decreasing $N_{qp}^\mathrm{slab}(t)$.

\begin{figure}
 \includegraphics[width=\columnwidth,totalheight=0.8\textheight,keepaspectratio]{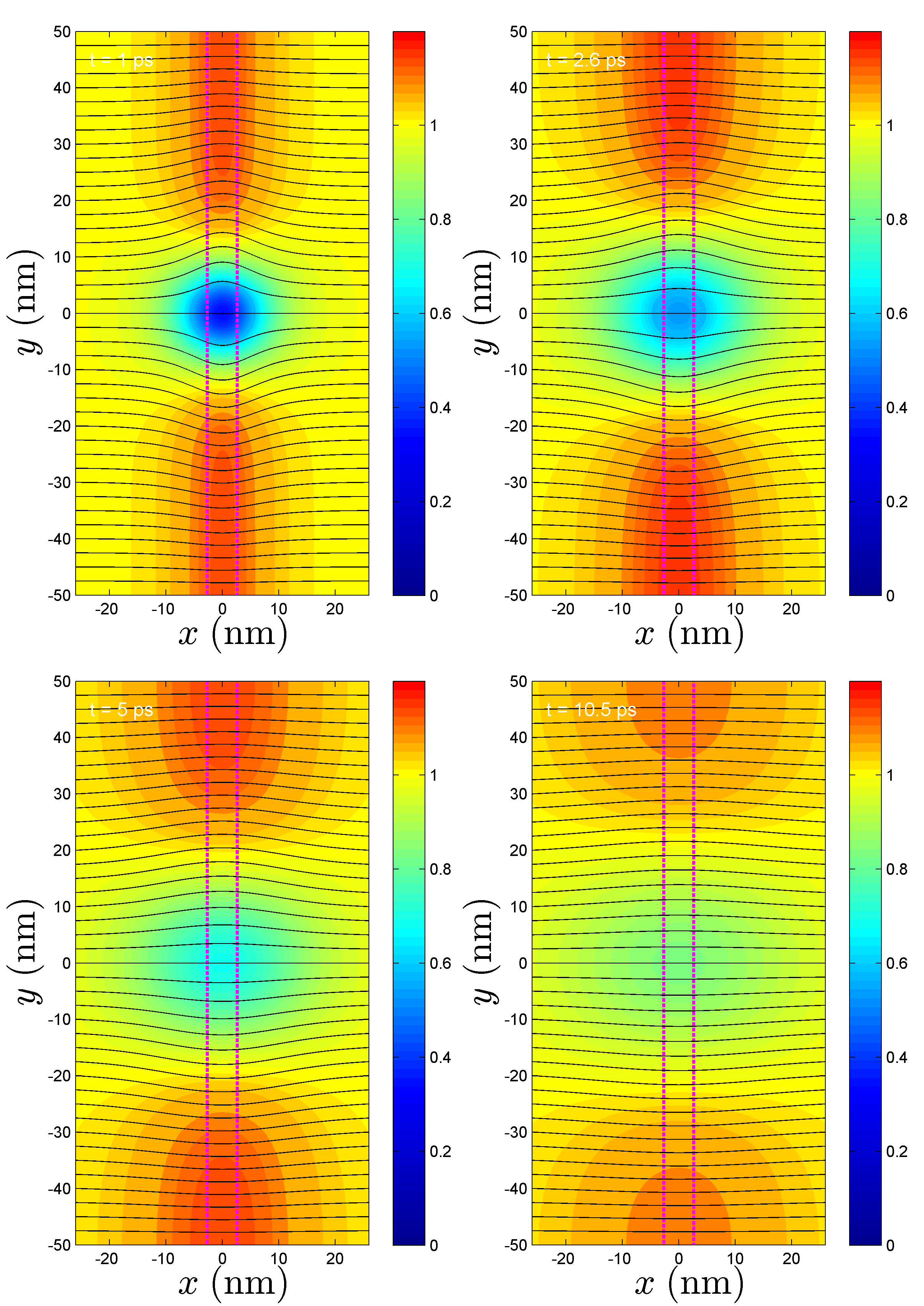}%
 \caption{Calculated relative current densities after the photon absorption, based on the QP densities shown in Fig.~\ref{Fig.QPdensity}. The vertical dashed lines outline the $\xi$-slab, and the black lines are streamlines of the bias current.\label{Fig.Currdensity}}
\end{figure}

\begin{figure}
 \includegraphics[width=\columnwidth,totalheight=200mm,keepaspectratio]{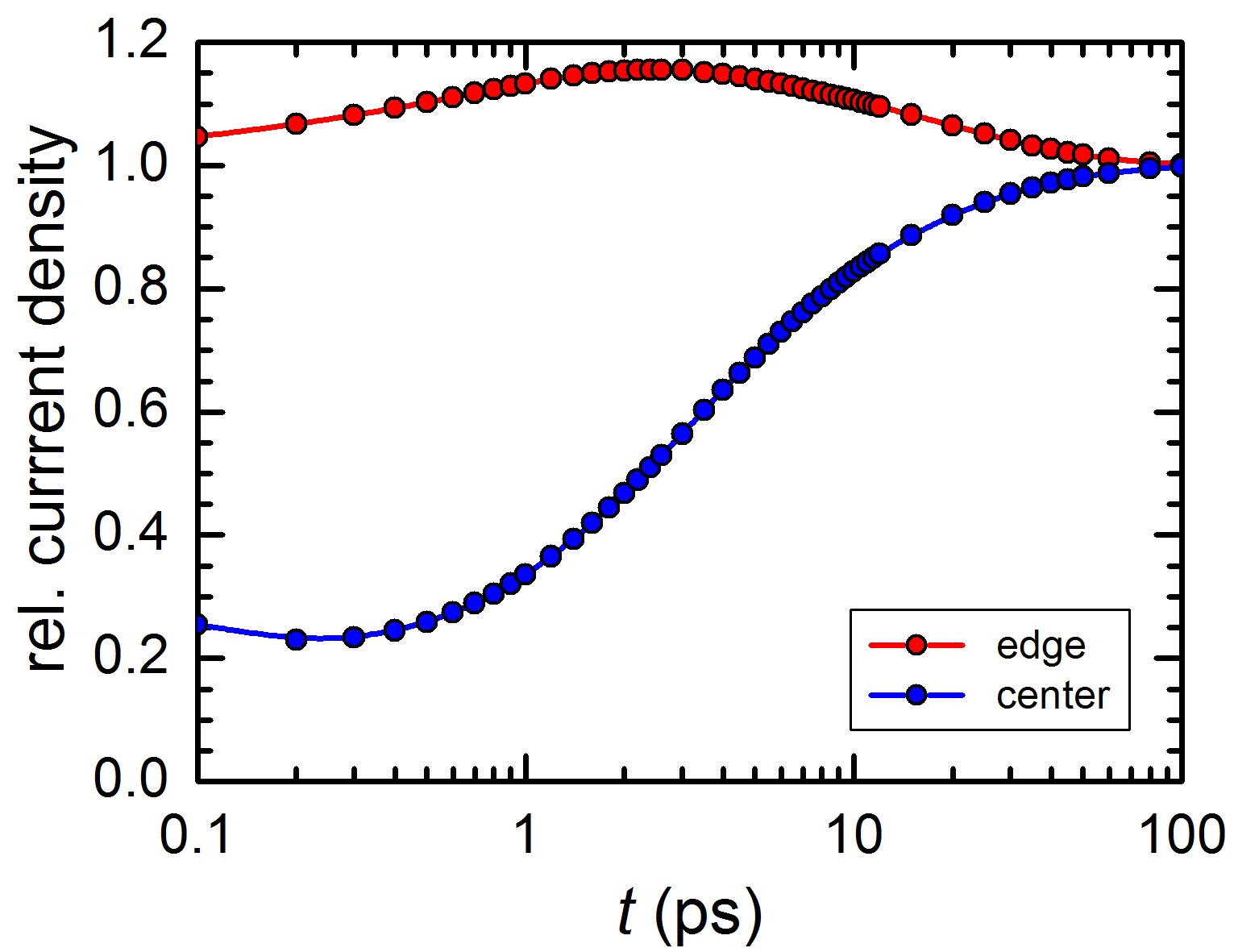}%
 \caption{Relative current densities as a function of time for two different positions extracted from the current distributions shown in Fig.~\ref{Fig.Currdensity}. The center position is the absorption site of the photon, and the edge position is one coherence length away from the geometric edge of the strip, at the same $x=0$ as the center position. The maximum current-density increase near the edge is realized at $t\approx2.6$ ps after the absorption of the photon. For clarity, the time axis is plotted on a logarithmic scale.\label{Fig.CurrentProfiles}}
\end{figure}

As outlined above in Sec.~\ref{Sec.model} we used the calculated distributions of QPs to obtain the distribution of the bias current by requiring superconducting phase coherence, current continuity and $\operatorname{div}\vec{j}=0$. The resulting relative current distributions are shown in Fig.~\ref{Fig.Currdensity} for the same conditions as the QP density in Fig.~\ref{Fig.QPdensity}. Although the current suppression in the center of the QP cloud is strongest right after absorption of the photon ($t\approx1$~ps, Fig.~\ref{Fig.Currdensity}(a)), the maximum current density in the side-walks is again reached for $t\approx t_{max}=2.6$~ps. This fact becomes even clearer in Fig.~\ref{Fig.CurrentProfiles}, where we plot the relative current densities as a function of time after the absorption of a photon for two different positions in the strip. The center position is the photon absorption site, and the edge position is one coherence length away from the geometrical edge of the strip. For the considered situation we obtained a maximum current increase of about $16$\% near the strip edges. Again, diffusion dominates over QP-multiplication for $t>t_{max}$, and the current distribution becomes more homogeneous as time progresses.

\begin{figure}
 \includegraphics[width=\columnwidth,totalheight=200mm,keepaspectratio]{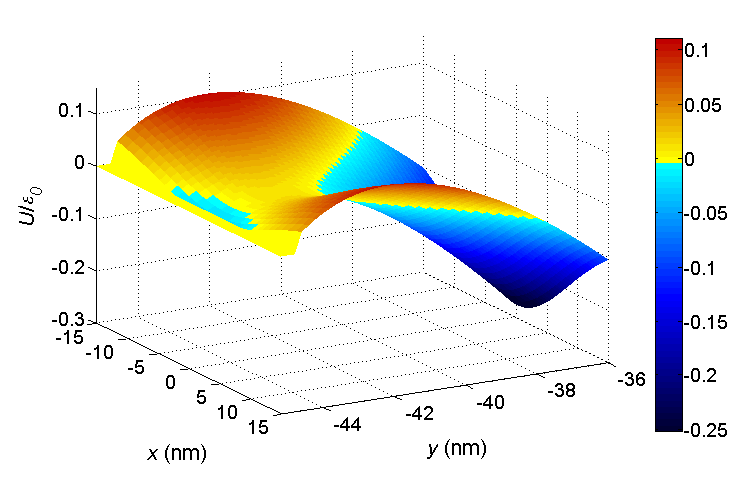}%
 \caption{Potential energy landscape for a vortex near the strip edge calculated for $t=2.6$~ps after photon absorption with an applied bias current $I/I_c=0.85$. The energy has been scaled by the vortex self-energy $\varepsilon_0$. The calculations have been done for the current distribution shown in Fig.~\ref{Fig.Currdensity}(b) caused by the absorption of a $1000$~nm photon at the center.\label{Fig.Barrier3D}}
\end{figure}

These inhomogeneous current distributions in turn affect the entry barrier for vortices at the edges. In Fig.~\ref{Fig.Barrier3D} we show a three dimensional representation of the potential energy landscape for a vortex near the edge of the strip. It has been calculated based on the current distribution in Fig.~\ref{Fig.Currdensity}(b) at $t=2.6$~ps after photon absorption and for an applied bias current $I/I_{c,v}=0.85$. For this particular situation the barrier remains positive and a vortex would still need additional thermal energy to enter the strip and trigger the formation of the initial normal conducting cross-section.

\begin{figure}
 \includegraphics[width=\columnwidth,totalheight=200mm,keepaspectratio]{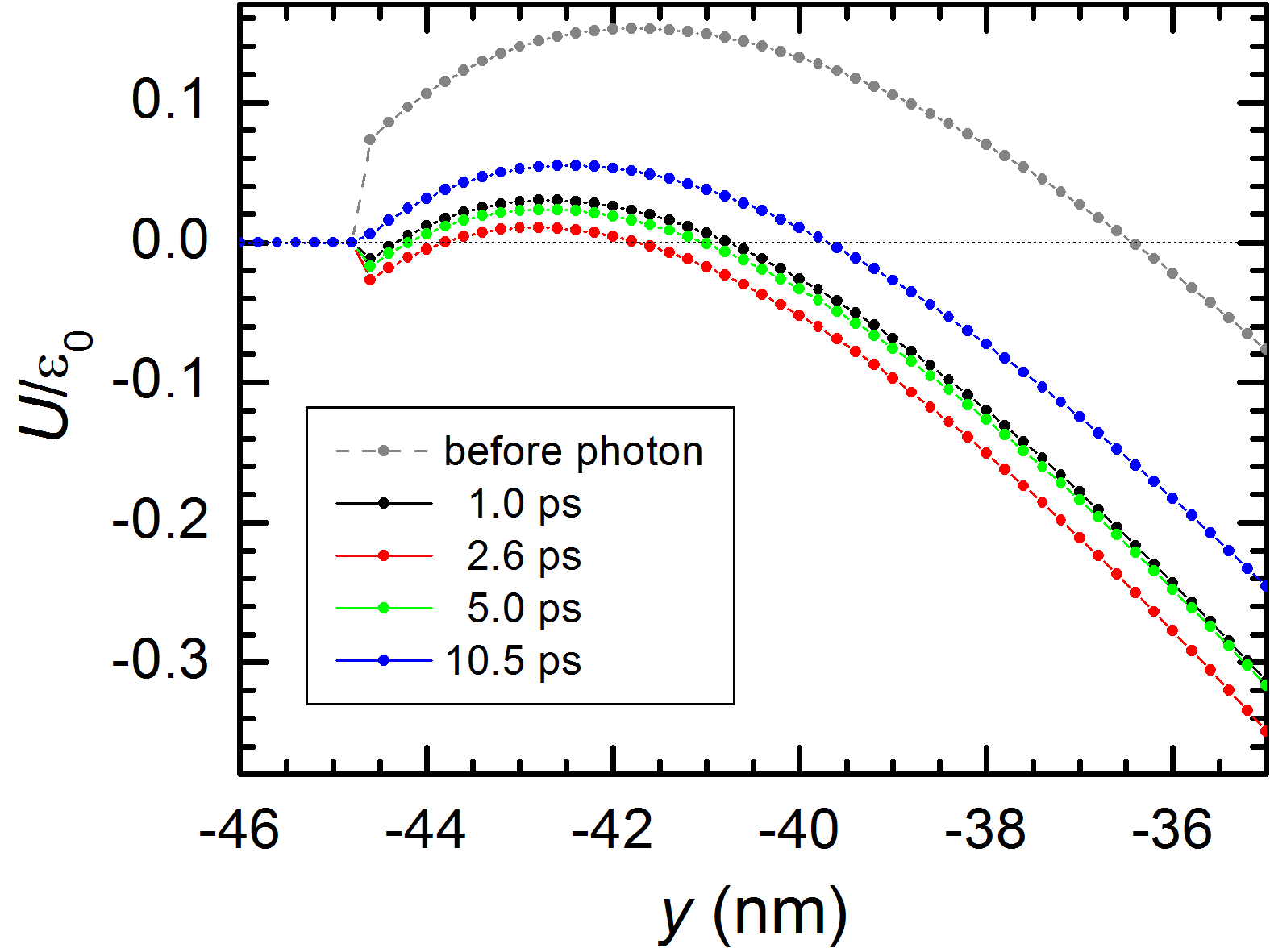}%
 \caption{Temporal evolution of the vortex-entry barrier along the cross-section containing the photon absorption site. The strongest influence of the excess QPs is again realized at $t\approx2.6$~ps after the photon absorption when the number of excess QPs in the $\xi$-slab reaches its maximum. For comparison the energy barrier in the undisturbed situation before photon absorption is also shown (gray dots).\label{Fig.Barrier}}
\end{figure}

In order to show the temporal evolution of the energy barrier, we plot in Fig.~\ref{Fig.Barrier} the potential energy for a vortex near the strip edge as a function of position across the strip for the cross section containing the photon absorption site at $y=0$. We observe again that the strongest reduction of the barrier occurs at around $t=2.6$~ps (red dots) after photon absorption when the number of excess QPs in the $\xi$-slab reaches its maximum. For comparison we have also plotted the energy barrier before the photon absorption (gray dots). The maximum of the curves in Fig.~\ref{Fig.Barrier} corresponds to the saddle-point of the 3D-energy landscape in Fig.~\ref{Fig.Barrier3D}. The criterion for a photon-detection event based on the vortex entry mechanism corresponds to a saddle-point value $\leq0$.

\subsection{Photon detection as functions of photon energy and bias current}

Next we evaluate the minimum (or threshold) bias-current $I_{th}$ for the detection of the photon in all three detection models for a given photon energy. First of all, it is important to realize that the critical current in the ``vortex model'' is the current for which the edge barrier vanishes, whereas in the other two models the relevant current scale is the depairing current. Within the vortex model developed in Refs.\ \onlinecite{Bulaevskii11} and \onlinecite{Bulaevskii12}, $I_{c,v}\approx0.826\Idep$. For the remainder, we will express all reduced bias currents scaled with the depairing critical current $\Ibias/\Idep$.

\begin{figure}
 \includegraphics[width=\columnwidth,totalheight=200mm,keepaspectratio]{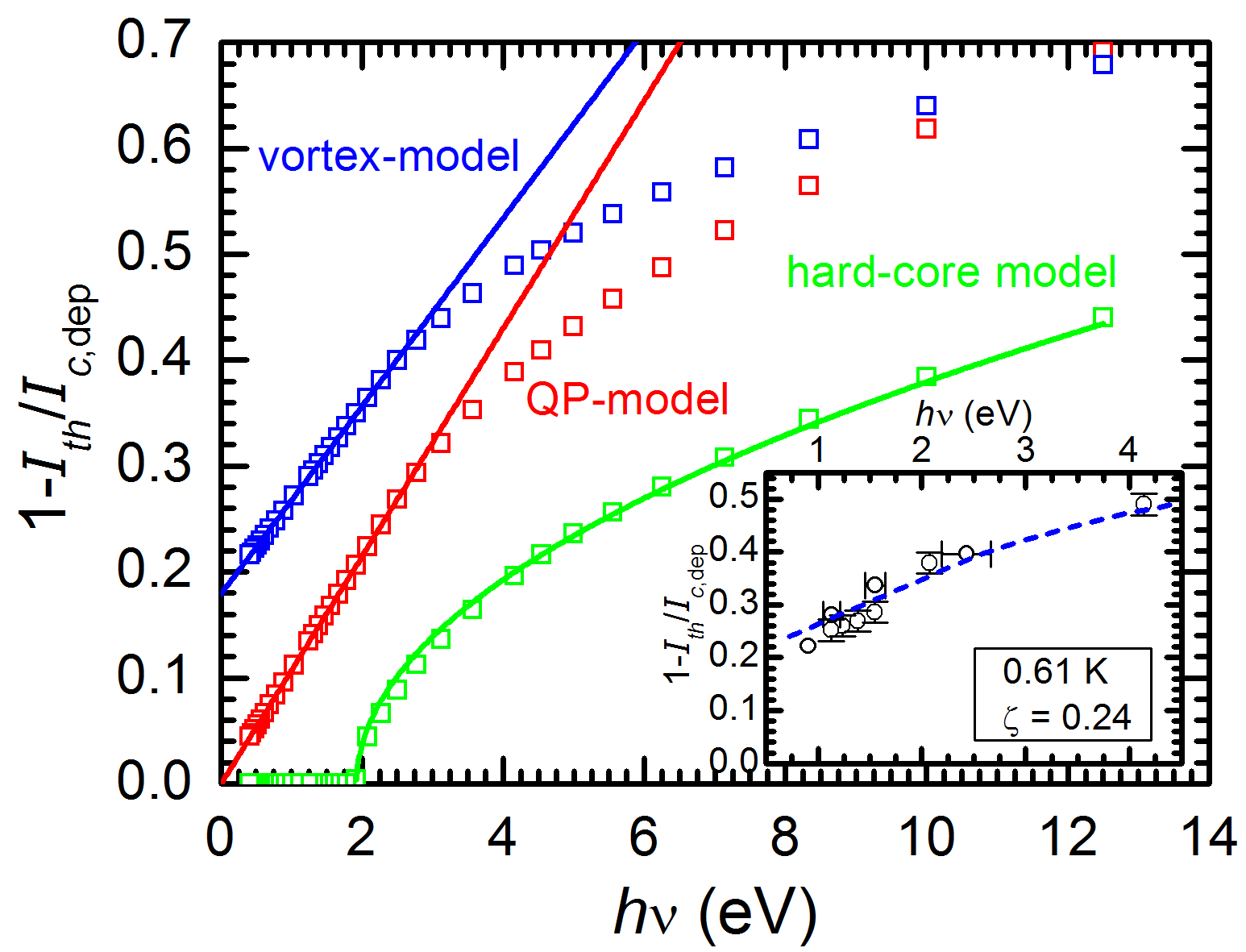}%
 \caption{Threshold current plotted as $1-I_{th}/\Idep$ \emph{vs.} the photon energy $h\nu$ for all three detection models as indicated. The temperature was set to $T=0.05\,T_c$. Solid lines are least-squares fits to the simulation data obtained for the hard-core and QP-model as explained in the text. For the QP-model only data points with $h\nu<1.9$\ eV have been considered for the fit. In the vortex model $I_{th}$ appears also to follow a linear dependence on $h\nu$ (blue line) up to photon energies that lead to the formation of a normal-conducting core. The inset shows the result of a least-squares fit of the vortex-model simulation data (blue dashed line) to experimental data from a TaN SNSPD (black circles) at $T=0.61~\text{K}\approx0.07\,T_c$ with the conversion efficiency $\zeta=0.24$ as the only adjustable parameter.\label{Fig.Threshold}}
\end{figure}

From the detection criterions in the hard-core model and QP model, Eqs.~\eqref{Eq.HardCoreCriterium} and \eqref{Eq.CriticalCurrentReduction}, respectively, one can derive explicit relations between photon energy and threshold current. In the latter model, the number of QPs in the $\xi$-slab $N_{qp}^\mathrm{slab}(t)$ is directly proportional to the photon energy, thus we expect $h\nu\propto1-I_{th}/\Idep$. In the hard-core model one usually assumes a cylindrical normal conducting volume $\pi \Rhc^2 d\propto h\nu$. Insertion in Eq.~\eqref{Eq.HardCoreCriterium} results in $\sqrt{h\nu}\propto1-I_{th}/\Idep$. In order to check these relations we plot in Fig.\ \ref{Fig.Threshold} the quantity $1-I_{th}/\Idep$ for all three models as a function of the photon energy up to energies $\approx 12.4$\ eV, corresponding to $\lambda=100$\ nm. The solid lines are least-squares fits to the corresponding data (green and red) confirming the expectations, namely a linear relationship for the QP model and a square-root behavior for the hard-core model. However, for high photon energies, $h\nu\gtrsim3$\ eV, the simulated threshold currents in the QP model deviate systematically from the expected linear behavior. The reason for this discrepancy is the appearance of a normal conducting core at these high photon energies, which is neglected in the derivation of the linear relation in that model.

For the vortex model we are not aware of any analytical approximation describing the threshold current dependence on photon energy, except for the gross approximation of a uniform density of excess QPs \cite{Lusche13a}. Our simulation results (blue data in Fig.\ \ref{Fig.Threshold}) suggest a similar linear dependence as for the QP-model at low photon energies, but shifted upwards along the $y$-axis as a consequence of the different current scales. Once the photon energy is high enough to allow for the formation of the normal conducting core, the simulated data deviate from the linear extrapolation in a similar way as in the QP-model.

Comparing the results of the three models with each other one can easily see that for a given bias current the vortex model requires the lowest photon energy and the hard-core model the highest energy. Photons with a fixed energy, on the other hand, are detected at the lowest bias currents in the vortex model, and at the highest current in the hard-core model. In fact, in the latter model photons may be absorbed without a detection event if the photon energy is too small, even in the hypothetical case of bias currents approaching the depairing critical current. With the simulation parameters given in Tab.~\ref{Tab.Material} and \ref{Tab.Geometry} the minimum detectable photon energy would be $h\nu\approx1.9$ eV corresponding to $\lambda\approx650$ nm. These numbers, however, strongly depend on the choice of the conversion efficiency $\zeta$.

Since in all three models the maximum number of excess QPs $\Nmax\propto\zeta h\nu$, our results obtained for the particular value $\zeta=0.25$ can be easily recalculated for any value of $\zeta$ by an appropriate re-scaling of the photon energy $h\nu$. This has been explicitly verified by running simulations with different values for $\zeta$. This allows us to directly compare our results with experimental data obtained for the threshold current as a function of photon energy. Corresponding data measured on a TaN SNSPD with very similar parameters as used in our calculations \cite{Engel12} are plotted in the inset of Fig.~\ref{Fig.Threshold}. These measurements were done at $T=0.61$~K~$\approx0.07\,T_c$. The original data were plotted as $1-I_{th}/I_c$, with $I_c$ the experimental critical current. We calculated the theoretical depairing critical current using the two-fluid temperature-dependence and the GL approximation (see Appendix \ref{App.T-dependence}) resulting in $I_c/\Idep\approx0.85$, and we re-scaled the experimental threshold currents accordingly. A satisfactory description of the experimental data is clearly only possible with the simulation results from the vortex model. Using the method of least-squares we fitted the calculated threshold currents to the experimental data by re-scaling the photon energy and obtain a conversion efficiency $\zeta\approx0.24$ with a conservative error estimate of $\pm0.04$. The corresponding best fit is shown as the blue dotted line in the inset of Fig.~\ref{Fig.Threshold}. The linear relation between threshold currents and excitation energy have also been found in experiments with a variant of an SNSPD\cite{Renema13a} over a much larger range of energies. The fact that a deviation from the linear behavior is not seen in Ref.~\onlinecite{Renema13a}, even at very high excitation energies, may be a consequence of multi-photon absorption instead of absorbing a single photon with the same energy.

Threshold currents as a function of photon energy have also been calculated with NbN material parameters. A comparison with the results for TaN reveals qualitatively the same dependence of $I_{th}$ on the photon energy, but shifted to higher current values or higher photon energies, respectively. In a typical experimental situation, in which one chooses a fixed bias current and determines the minimum energy for direct photon detection, this energy turns out to be a factor $\approx 2.5$ larger in all there detection models for using NbN as detector material as compared to TaN. This comparison alone does not allow to distinguish between the considered detection models, but it confirms the experimentally observed differences between the cut-off wavelengths of TaN and NbN SNSPDs \cite{Engel12}.

As mentioned in the introduction, it is not clear at present, how to correctly treat the vortex potential energy when the vortex resides within $\approx\xi$ of the strip edge. At the current stage this leaves some uncertainty, for example, with respect to the critical current for a vanishing vortex barrier. In terms of the photon detection, an improved vortex model could result in a small shift up or down of the blue data points in Fig.~\ref{Fig.Threshold}, but we do not expect any fundamental change of the results presented here.

\subsection{Trigger times as functions of photon energy}

\begin{figure}
 \includegraphics[width=\columnwidth,totalheight=200mm,keepaspectratio]{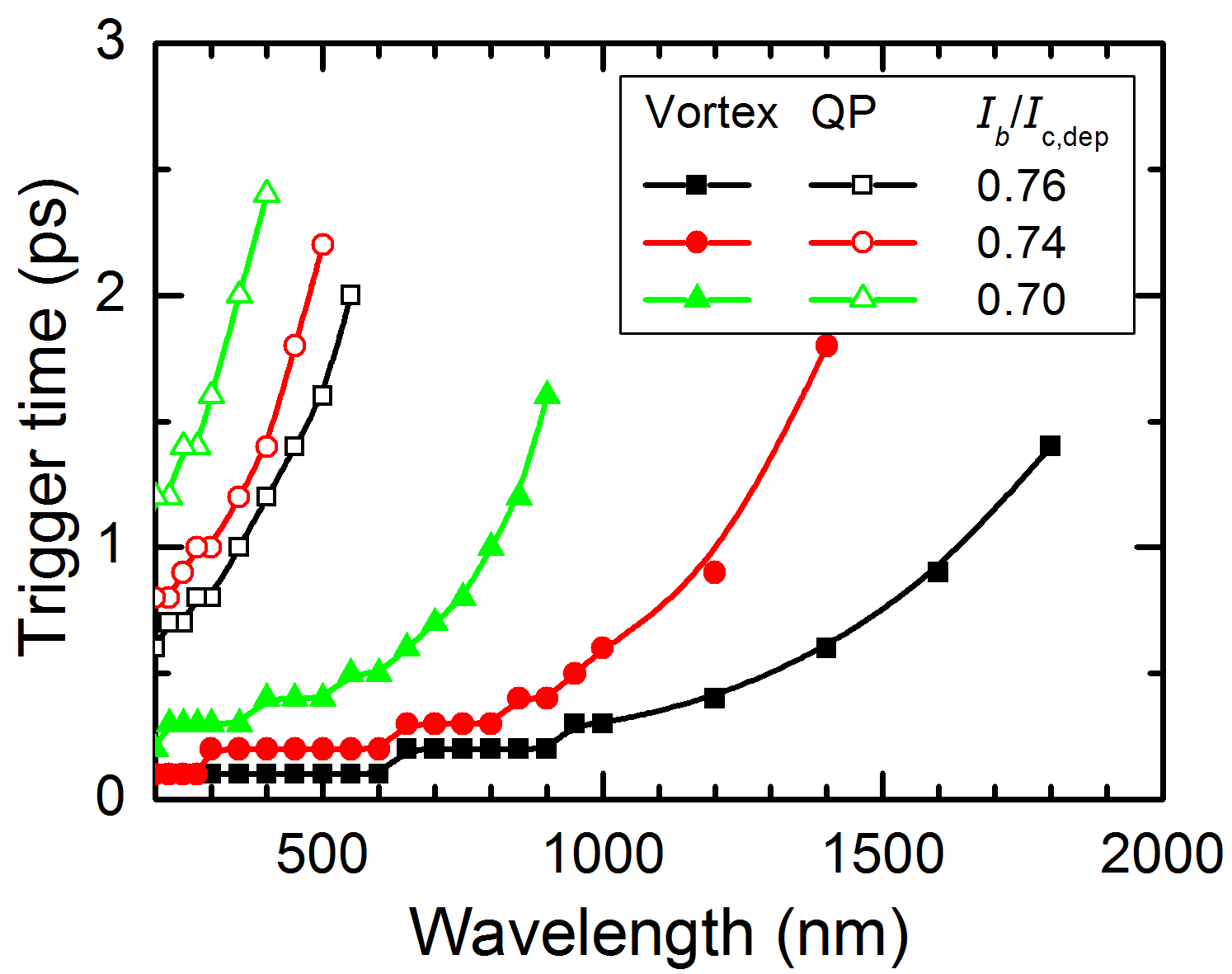}%
 \caption{Time delay between absorption of a photon and trigger of the initial normal-conducting cross-section as a function of photon wavelength (lines are guides, only). Shown are simulation results for three different bias currents. The temperature was set to $T=0.05\,T_c$. The vortex mechanism always leads to a detection event well before the QP-model. For the chosen parameters, no detection event would be registered in the hard-core model.\label{Fig.Trigger}}
\end{figure}

For certain combinations of bias current and photon energy, two or all three models allow for the direct detection of the photon. For $1-I_b/\Idep=0.2$ and $h\nu=3.13$\ eV ($400$\ nm), for example, the vortex and the QP-model result in the trigger of the initial normal conducting cross-section, but not necessarily at the same time after photon absorption. For a given bias current we evaluate the spatial QP-distributions as functions of time and wavelength and determine the instant $\tau_\mathrm{trigger}$ when the detection criterion is fulfilled in the three detection models. The results are plotted in Fig.~\ref{Fig.Trigger} for three different bias currents $I_b/\Idep=0.70,\ 0.74\ \text{and}\ 0.76$, corresponding to $I_b/I_{c,v}=0.85,\ 0.90\ \text{and}\ 0.925$, for the vortex (filled symbols) and QP-models (open symbols). Results for the hard-core model are not shown in Fig.~\ref{Fig.Trigger}, because at the conversion efficiency $\zeta=0.25$ and the shortest considered photon wavelength $\lambda=300$\ nm even higher bias-currents $I_b/\Idep>0.8$ or $I_b/I_{c,v}>0.97$ would be required to fulfill the detection criterion. Even if one assumes bias currents so close to $I_{c,v}$, it turns out that the detection criterion in the hard-core model will be reached for even longer delays after photon absorption than in the QP-model.

In the vortex model, at these high bias currents relative to the critical current for a vanishing vortex-entry barrier shown in Fig.~\ref{Fig.Trigger}, only a very small number of excess QPs is necessary to alter the current distribution sufficiently to suppress the remaining energy barrier. Accordingly, the detection criterion in the vortex model is reached almost immediately after photon absorption. For longer wavelengths closer to the cut-off wavelength, the time delay increases towards $\approx 2.6$~ps, the time at which the maximum suppression of the energy barrier is expected. The trigger times in the vortex model are all smaller than $2.6$~ps in Fig.~\ref{Fig.Trigger}, because of the fixed values of $\lambda$. For all currents and wavelengths investigated, the trigger times in the QP-model are significantly longer than in the vortex model. We may therefore conclude that the detection via the entry of a magnetic vortex is the primary mechanism to trigger the formation of a normal-conducting cross section, not only as a function of photon energy and bias current, but also temporally in situations when the other detection models might allow for a direct, but delayed, detection event.

\section{Conclusions}

\begin{table}
\caption{\label{Tab.Summary}Comparison of the results for the different detection models. Note on $\lim_{h\nu\rightarrow0} I_{th}$: The minimum photon energy that can possibly be detected equals to $2\Delta$.}
\begin{tabular}{lllc}
model & $I_{th}$ & $\lim\limits_{h\nu\rightarrow0} I_{th}$ & $\tau_\mathrm{trigger}$\\
\hline
hard-core & $\propto\sqrt{h\nu-E_0}$ & not detected & $\tau_{hc}>\tau_{qp}>\tau_{v}$\\
QP & $\propto h\nu$ & $=\Idep$ & $\tau_v<\tau_{qp}\lesssim2.6$ ps\\
vortex & $\propto h\nu$ & $=I_{c,v}<\Idep$ & $\lesssim2.6$ ps
\end{tabular}
\end{table}

We have presented a simple numerical model based on the diffusion of QPs generated after photon absorption in a thin superconducting film. The results were applied to predict the detection of an absorption event in SNSPD by comparing three currently considered detection mechanisms. All our results summarized in Table \ref{Tab.Summary} are in favor of a vortex assisted detection mechanism, whereby the excess QPs lead to the suppression of the edge barrier for vortex entry and the subsequent dissipative crossing of a vortex. This process triggers the initial normal conducting domain in the superconducting strip. Competing mechanisms require higher photon energies and occur at a later stage of the QP multiplication and diffusion process.

We also compared our numerical results with experimental data. We obtain good agreement for the dependence of the threshold energy on the bias current and the different cut-off wavelengths for NbN and TaN SNSPD.

At the current stage our numerical model is by no means complete. Its aim is to capture the most important processes in the detection event of a photon in SNSPD and it still contains a number of simplifications and assumptions. Despite this simplicity it allows us to obtain a better understanding of the first stages of the detection process in SNSPD, which are otherwise difficult to probe experimentally. Further refinements of this model may allow for a better understanding of additional aspects of the detection process and for the development of a more complete theoretical description.

\section*{Acknowledgement}

This research received support from the Swiss National Science Foundation grant No.\ 200021\_135504/1 and 200021\_146887/1. We thank A.~Semenov, J.~Renema and D.~Y.~Vodolazov for fruitful discussions.

\appendix

\section{Analytic solution of partial differential equations\label{App.Analytic}}

The set of partial differential equations \eqref{Eq.PDE_e} and \eqref{Eq.PDE_QP} can be solved analytically for the case of a two-dimensional film and making the assumption $D_e=D_{qp}=D$. We then have
\begin{widetext}
\begin{align}
\frac{\partial C_e(\vec{r},t)}{\partial t} &= D\nabla^2 C_e(\vec{r},t),\label{Eq.App.PDE_e}\\
\frac{\partial C_{qp}(\vec{r},t)}{\partial t} &= D\nabla^2 C_{qp}(\vec{r},t) - \frac{C_{qp}(\vec{r},t)}{\tau_r} + \frac{\zeta h\nu}{\Delta\tau_{qp}}\exp\left(-\frac{t}{\tau_{qp}}\right)C_e(\vec{r},t).\label{Eq.App.PDE_qp}
\end{align}
Eq.~\eqref{Eq.App.PDE_e} has the well-known solution Eq.~\eqref{Eq.2D_Diffusion}
\begin{equation}
C_e(r,t) = \frac{1}{4\pi Dt}\exp\left(-\frac{r^2}{4Dt}\right),
\end{equation}
and inserted into \eqref{Eq.App.PDE_qp} results in
\begin{equation}
\frac{\partial C_{qp}(\vec{r},t)}{\partial t} = D\nabla^2 C_{qp}(\vec{r},t) - \frac{C_{qp}(\vec{r},t)}{\tau_r} + \frac{\zeta h\nu}{\Delta\tau_{qp}}\frac{1}{4\pi Dt}\exp\left(-\frac{t}{\tau_{qp}}\right)\exp\left(-\frac{r^2}{4Dt}\right),\label{Eq.App.inhomogeneous}
\end{equation}
\end{widetext}
which is an inhomogeneous differential equation. It can be analytically solved yielding Eq.~\eqref{Eq.AnalyticalSolution}.

\section{Temperature dependence of material parameters\label{App.T-dependence}}

Calculations were done at a very low temperature $T\ll T_c$ to eliminate or at least reduce thermal effects. However, we could not assume $T=0$, because the diffusion coefficient of QPs $D_{qp}=0$ for $T\rightarrow0$. We calculated values for $T$-dependent parameters using the following expressions and methods.

The BCS $T$-dependence of the superconducting gap can be well approximated by the simple formula
\begin{equation}
\Delta(\tau) = \Delta\left(1-\tau^2\right)^{0.5}\left(1+\tau^2\right)^{0.3},\label{Eq.App.Delta}
\end{equation}
with $\tau=T/T_c$ the reduced temperature and $\Delta=\alpha k_BT_c$, where the BCS-value\cite{Tinkham96} $\alpha=1.764$ has been used for TaN and $\alpha=2$ for NbN\cite{Romestain04,Henrich12a}.

For the coherence length $\xi(\tau)$ we use an interpolation formula \cite{Bartolf10}
\begin{equation}
\xi(\tau) = \xi\left(1-\tau\right)^{-0.5}\left(1+\tau\right)^{-0.25},\label{Eq.App.coherence}
\end{equation}
with $\xi=\sqrt[4]{2}\xi_{GL}$ and $\xi_{GL}$ the extrapolated GL-coherence length at $T=0$. Eq.~\eqref{Eq.App.coherence} smoothly interpolates between the GL-result near $T_c$ and the estimated zero-temperature value for a dirty type-II superconductor at $T=0$\cite{Werthamer66}.

The magnetic penetration depth in the dirty limit is given by \cite{Tinkham96}
\begin{equation}
\lambda(\tau) = \lGL\left(\frac{\Delta(\tau)}{\Delta}\tanh\left[\frac{\Delta(\tau)}{2k_B\tau T_c}\right]\right)^{-0.5},
\end{equation}
with $k_B$ the Boltzman constant.

The theoretical depairing critical current is calculated using the two-fluid temperature dependence and the GL approximation \onlinecite{Bulaevskii11}
\begin{equation}
\Idep=\frac{\Phi_0 wd}{3\pi\sqrt{3}\mu_0\lGL^2\xi}\left(1-\tau^2\right)\left(1-\tau^4\right)^{0.5}.
\end{equation}

A simple analytical formula for the temperature dependence of the normalized diffusion coefficient for QPs in the superconducting state can be derived in the limiting case of $T\ll T_c$ ($\tau\lesssim0.1$) that resembles the diffusion of an ideal gas. For the thermal conductivity we use the approximation \cite{Abrikosov88}
\begin{equation}
\kappa_s \approx 2 N_0DT\left(\frac{\Delta}{k_B T}\right)^2\exp\left(-\frac{\Delta}{k_B T}\right),
\end{equation}
which, divided by $\kappa_n(T_c)$, results in
\begin{equation}
\tilde{\kappa}(\tau) \approx \left(\frac{\Delta}{k_B T_c}\right)^2\frac{6}{\pi^2\tau}\exp\left(-\frac{\Delta}{k_B T_c\tau}\right).
\end{equation}

An expression for the normalized specific heat at $T\ll T_c$ has already been given in Ref.\ \onlinecite{Bardeen57},
\begin{widetext}
\begin{equation}
\tilde{C}(\tau)=\frac{C_s(T)}{C_n(T_c)}\approx \frac{6}{\pi^2}\sqrt{\frac{\pi}{2}}\left(\frac{\Delta}{k_B T_c}\right)^{2.5}\tau^{-1.5}\exp\left(-\frac{\Delta}{k_B T_c\tau}\right),
\end{equation}
\end{widetext}
where we have already replaced the modified Bessel-functions by their first order approximation for large arguments $\frac{\Delta}{k_B T_c\tau}\gg1$. It is then easy to see that
\begin{equation}
\frac{\tilde{\kappa}(\tau)}{\tilde{C}(\tau)} = \frac{D_{qp}(\tau)}{D_e(T_c)} = \sqrt{\frac{2}{1.764\pi}}\sqrt{\tau}\approx 0.6\sqrt{\tau}.\label{Eq.App.lowTApprox}
\end{equation}

\section{Calculation of single-vortex potential for inhomogeneous current distribution\label{App.VortexPotential}}

Eq.~\eqref{Eq.VortexPotential} is the potential energy of a single-vortex inside a current-carrying superconducting strip with $d\ll w\ll\Lambda$ and a homogeneous current density. One may calculate the force on a vortex inside the strip, $\vec{F}=-\vec{\nabla}U$, which by symmetry has only a component in the $y$-direction,
\begin{equation}
F_y(y) = \varepsilon_0\left[\frac{\pi}{w}\tan\left(\frac{\pi y}{w}\right) + \frac{2w j_x}{I_{c,v}\exp(1)\xi}\right],\label{Eq.App.Force_homo}
\end{equation}
with $j_x=I/w$ the thin-film current-density in the $x$-direction. The first term on the right-hand side of \eqref{Eq.App.Force_homo} is the Lorentz-force $j\Phi_0$ on a vortex caused by the current-density from the chain of image vortices and antivortices, necessary to fulfill the boundary condition at the strip edges, namely that the current density can only have a longitudinal component at the edges. The second term is the Lorentz-force exerted by the applied bias current.

After absorption of a photon and the creation of the diffusing QP-cloud, the current density becomes a function of $x$ and $y$, $j_x(x,y)$, and the energy scale $\varepsilon_0$ becomes also a function of vortex position. It may be expressed as
\begin{equation}
\varepsilon_0(x,y) = \frac{\Phi_0^2}{2\pi\mu_0\Lambda} = \frac{\Phi_0^2 e^2n_{se}^{2D}(x,y)}{4\pi m_e}\label{Eq.App.SelfEnergy},
\end{equation}
where we have used the definition of the London penetration depth in the last expression, with $\left|e\right|$ the elementary charge and $m_e$ the electron mass. Replacing the $j_x$ and $\varepsilon_0$ by their position dependent counterparts in Eq.~\eqref{Eq.App.Force_homo} and dividing it by the equilibrium value $\varepsilon_0$, the force on a vortex becomes
\begin{widetext}
\begin{equation}
\frac{F_y(x,y)}{\varepsilon_0} = \frac{n_{se}^{2D}(x,y)}{n_{se}^{2D}}\left[\frac{\pi}{w}\tan\left(\frac{\pi y}{w}\right) + \frac{2w j_x(x,y)}{I_{c,v}\exp(1)\xi}\right].\label{Eq.App.Force_inhom}
\end{equation}
The potential energy $U(x,y)/\varepsilon_0$ can then be calculated from \eqref{Eq.App.Force_inhom} by numerical integration over $y$ for fixed $x$:
\begin{equation}
\frac{U(x,y)}{\varepsilon_0} = \frac{\pi}{w}\int_{\frac{\xi-w}{2}}^{\frac{w-\xi}{2}}\frac{n_{se}^{2D}(x,y)}{n_{se}^{2D}}\tan\left(\frac{\pi y}{w}\right)\mathrm{d}y + \frac{2w}{I_{c,v}\exp(1)\xi}\int_{-\frac{w}{2}}^{\frac{w}{2}}\frac{n_{se}^{2D}(x,y)}{n_{se}^{2D}}j_x(x,y)\mathrm{d}y,
\end{equation}
where the integration limits have been chosen to obtain the same normalization of $U(x,y)$ as used in Ref.\ \onlinecite{Bulaevskii12}.
\end{widetext}

\bibliography{Literature}

\end{document}